\documentclass[twocolumn,amssymb,amsmath,aps,prb]{revtex4}
\usepackage[dvips]{graphicx}
\begin{document}
\title{Two dimensional magnetic correlation in the unconventional corrugated layered oxides (Ba,Sr)$_4$Mn$_3$O$_{10}$}
\author{J. Sannigrahi}
\author{S. Chattopadhyay}
\altaffiliation{Present address: Laboratoire de Physique des Solides, Universit\'e Paris-Sud, France}
\author{A. Bhattacharyya}
\altaffiliation{Present address: ISIS Facility, Rutherford Appleton Laboratory, UK}
\author{S. Giri}
\author{S. Majumdar}
\email{sspsm2@iacs.res.in} 
\affiliation{Department of Solid State Physics, Indian Association for the Cultivation of Science, 2A \& B Raja S. C. Mullick Road, Jadavpur, Kolkata 700 032, India }
\author{D. Venkateshwarlu}
\author{V. Ganesan}
\affiliation{UGC-DAE Consortium for Scientific Research,University Campus, Khandwa Road, Indore 452 017, India}
\begin{abstract}
Both Ba$_4$Mn$_3$O$_{10}$ and Sr$_4$Mn$_3$O$_{10}$ crystallize in an orthorhombic crystal structure consisting of corrugated layers containing 
Mn$_3$O$_{12}$ polydedra. The thermal variation of magnetic susceptibility of the compositions consists of a broad hump like feature indicating the presence of low dimensional magnetic correlation. We have systematically investigated the magnetic data of these compounds and found that the experimental results  match quite well with the two dimensional Heisenberg model of spin-spin interaction. The two dimensional nature of the magnetic  spin-spin interaction is  supported by the low temperature heat capacity data of Ba$_4$Mn$_3$O$_{10}$. Interestingly, both the samples show dielectric anomaly near the magnetic ordering temperature indicating multiferroic behavior.   

\end{abstract} 
\maketitle

\section{Introduction}
AMnO$_3$-type (A = alkaline or rare-earth metal) manganese oxides with Perovskite structure  attracted considerable attention primarily due to the observed colossal magnetoresistance (CMR) behavior.~\cite{salamon} Apart from CMR oxides, a wide variety of Mn-based ceramics exist in nature with complex crystallographic structure. Actually, the Perovskite based manganites belong to a wider class of materials with general formula A$_{n+1}$Mn$_n$O$_{3n+1}$, known as Ruddlesden-Popper phase (RP). The $n$ = 3  compound  is reported to form with A = Ca  with usual RP structure,~\cite{camno} which can be considered to have staking of trilayers made out of MnO$_6$ octahedra along the crystallographic $c$ axis. However, the Sr and Ba  counterparts, namely Sr$_4$Mn$_3$O$_{10}$ and Ba$_4$Mn$_3$O$_{10}$ respectively ~\cite{srmno, bamno}  do not crystallize in a tri-layered structure. For them  the crystal structure can be visualized as being built up from Mn$_3$O$_{12}$ trimers of face sharing octahedra.~\cite{srmno} These Mn$_3$O$_{12}$ trimers form two dimensional layers perpendicular to the $b$ axis of the crystal. The layers are separated from each other by Ba/Sr cations. Considering such  coordination of Mn atoms in these compounds, it is expected that the magnetism can be quite fascinating with primarily  three kinds of magnetic interactions: (i)intra-trimer, which operates between the Mn$^{3+}$ ions within a single Mn$_3$O$_{12}$ polyhedron; (ii) intra-layer, which is the magnetic interaction constrained in the two-dimensional (2D) layer; and (iii)  finally the inter-layer interaction which can lead to a three-dimensional (3D) magnetic ordering. 
\par
Magnetic studies indicate that both Ba$_4$Mn$_3$O$_{10}$ and Sr$_4$Mn$_3$O$_{10}$ undergo long range antiferromagnetic (AFM) ordering below the N\'eel temperature ($T_N$) close to 80 K.~\cite{tang,kb} However, precursor to the AFM ordering, both the samples show a broad hump like feature in the magnetization ($M$) versus temperature ($T$) data. Neutron data rule out the existence of any long range magnetic order associated with the hump like feature.~\cite{bamno}, however, long range magnetic order is quite clearly visible in the neutron data recorded at 5 K in BMO. The  prominent hump-like feature in the $M(T)$ data indicates that some sort of short range magnetic correlation is prevailing in the sample prior to the long range order. 
\par
Tang {\it et al.}~\cite{tang} fitted the hump like feature to Bonner-Fisher model and obtained the spin-spin coupling parameter to be $J$= 200 K for Sr$_4$Mn$_3$O$_{10}$. However, their fitting to the susceptibility data is not satisfactory and they obtained an effective magnetic moment (6.4 $\mu_B$/Mn)  which is much larger than the expected moment for $S$ = 3/2 state  (3.87 $\mu_B$/Mn). In addition, Bonner-Fisher model is supposed to be applicable to linear chain systems only,~\cite{bf} whereas the present composition has magnetic correlation constrained in a plane. There is so far no report on the analysis of high-$T$ hump in case of Ba-counterpart.

\par
It appears that the exact magnetic characterizations of Ba$_4$Mn$_3$O$_{10}$ and Sr$_4$Mn$_3$O$_{10}$ compounds are lacking in the literature.  Considering this, we performed a comprehensive investigation on Ba$_4$Mn$_3$O$_{10}$ (BMO) and as well as on Sr$_4$Mn$_3$O$_{10}$ (SMO). Our results indicate the existence of 2D magnetic correlation in the samples which is likely to be  associated with the layered structure of the compositions.       

\begin{figure}[t]
\centering
\includegraphics[width = 8.5 cm]{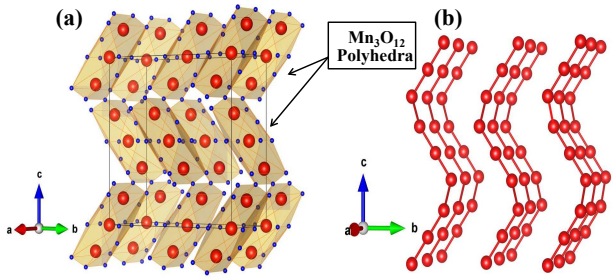}
\caption {(a) A perspective view of the crystal structure of BMO or SMO where Mn$_3$O$_{12}$ polyhedra are  indicated. The unit cell of the lattice is indicated by solid line. Here Mn and O are shown by big red sphere and small blue sphere respectively. (b) indicates few corrugated layers containing Mn$^{4+}$ ions in the $bc$ plane of the lattice, where only Mn atoms are shown for clarity.}
\label{structure}
\end{figure}

\section{Experimental Details and Crystal Structure}
Polycrystalline samples of BMO and SMO were prepared following solid state reaction route in air. Stoichiometric mixture of BaCO$_3$/SrCO$_3$ and MnO$_2$ were initially heated at $1000^\circ$ C for 48 h. The final sintering was performed at 1350$^\circ$ C for 3 days in pelletized form and then quenched to room temperature.~\cite{boulahya}
\par
Room temperature powder x-ray diffraction (XRD) measurement was carried out in a BRUKER diffractometer using Cu K$\alpha$ radiation. Rietveld refinements~\cite{hvr} were performed on the XRD patterns of both BMO and SMO using MAUD software package~\cite{maud} which show that the samples have orthorhombic unit cell with $Cmca$ space group. The reliability factor of the refinement is close to 1.42. The orthorhombic lattice parameters ($a$, $b$ and $c$) are all found to be slightly lower in case of SMO than BMO (table I). A perspective view of the unit cell of BMO/SMO is shown in fig. 1 (a). Mn$_3$O$_{12}$ polyhedra, built up from face-sharing MnO$_6$ octahedra, are   linked through common vertices to form corrugated layered structure perpendicular to the $b$ axis. Ba/Sr atoms occupy the space between the layers. 
\par
The magnetic measurements were performed on a Quantum Design SQUID magnetometer (MPMS-4, Evercool). The ac dielectric measurements were performed using an Agilent E4980A precision LCR meter in the temperature range 10-300 K in a helium closed cycle refrigerator. Electric polarization of the samples was calculated from the pyroelectric current measurement using a Kithley electrometer. Heat capacity ($C$) measurement was carried out using a Quantum Design physical properties measurement system.

\begin{table}[t]
\centering
\caption{The table depicts orthorhombic lattice parameters ($a$, $b$ and $c$), Mn-Mn nearest neighbour separation on a layer, intra-layer Heienberg interaction term ($J_{2D}$) as obtained from fitting the $\chi(T)$ data to eqn.~\ref{htse}, intra-layer interaction term as obtained by fitting to eqn.~\ref{mean_field} and antiferromagnetic ordering temperature ($T_N$).}
\centering
\begin{tabular}{|c|c|c|}
\hline
\hline
 & Ba$_4$Mn$_3$O$_{10}$ &  Sr$_4$Mn$_3$O$_{10}$ \\ 
\hline
Lattice parameters (\AA) &  $a$ = 5.677  & $a$ = 5.478,  \\
&$b$ = 13.106 & $b$ = 12.556 \\
&$c$ = 12.694  &  $c$ = 12.525 \\
\hline
Mn$_1$-Mn$_2$ distance (\AA)  &  2.57   & 2.51\\
\hline
$J_{2D} $ (K) & -45.5 (3)    & -46.9 (4) \\
\hline
$J^{\prime}$ (K) & -2.4(1)    & -2.5(1) \\
\hline
$T_N$ (K) & 80.0 & 82.4\\

\hline
\hline

\end{tabular}
\label{values}
\end{table}    

\begin{figure}[t]
\centering
\includegraphics[width = 8 cm]{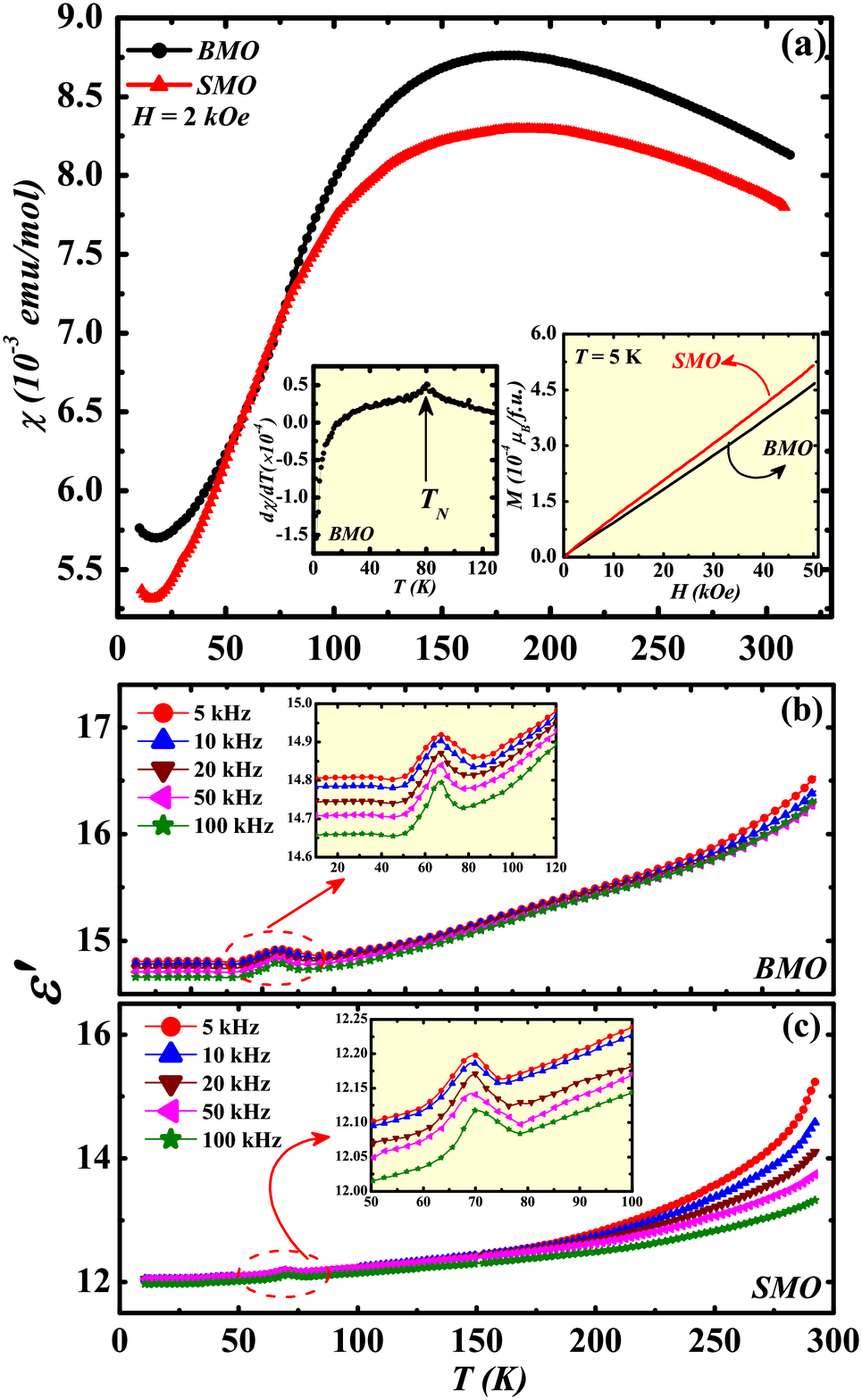}
\caption {(a) shows the temperature variation of dc magnetic susceptibility of both  BMO and SMO. The left inset shows the first order temperature derivative of $\chi(T)$ where the antiferromagnetic ordering temperature $T_N$ is indicated for BMO. The right inset shows the isothermal magnetization as a function of magnetic field for BMO and SMO recorded at 5 K. (b) and (c) respectively show the ac dielectric response of BMO and SMO measured at different frequencies  with inset in each panel depicting an enlarged view of the anomaly observed close to $T_N$.}
\label{MT}
\end{figure}

\section{Results}
Figure 2(a) shows the $T$ variation of dc magnetic susceptibility ($\chi = M/H$) of BMO and SMO measured at $H = 2$ kOe between 2 and 320 K. A broad hump is  noticed in the $\chi(T)$ for both compounds at par with the previous reports.~\cite{tang,bamno,kb}  The values of $T_{max}$ (the temperature at which maximum $\chi$ is observed) are found to be 179 K and 183 K  for BMO and SMO respectively. We  observed  anomalies (in the form of change in slope) at around   80 K ($\approx T_N$) for both the samples, which had been previously found to be the onset point of long range magnetic ordering from powder neutron diffraction and heat capacity measurements.~\cite{kb, bamno, tang} This anomaly is distinctly observed in the $\left({d\chi}/{dT}\right)$ data (left inset of  figure 2(a)) plotted for BMO. We observe a rise in the $\chi(T)$ data below about 10 K. Such low-$T$ rise in low-D magnetic sytems is generally referred as Curie tail.~\cite{chattopadhyay} Here the  extent of Curie tail is much smaller than the previously reported data~\cite{tang,bamno} indicating the lower concentration of  paramagnetic impurity in the presently studied samples. The right  inset of figure 2(a) shows the isothermal $M-H$ curves for the two compounds at 5 K. $M(H)$ data show almost linear $H$ dependence as expected from antiferromagnetically correlated spins. There are no sign of hysteresis in the isothermal magnetization data.
\par
Figure 2(b) and (c)  show the $T$ dependence of the real part of the complex dielectric permittivity ($\epsilon^{\prime}$) for BMO and SMO respectively measured at different ac frequencies. A small kink is observed near $T_N$ for both the compounds. We measured the electric polarization ($P$) of the samples through the measurement of pyroelectric current. However, we failed to observe any net $P$ below $T_N$ in case of these two samples, which rules out the possible ferroelectric transition at $T_N$. A thermally activated rise in $\epsilon^{\prime}$ is noted with the increase of temperature above 150 K in case of BMO and above 200 K for SMO. Such behaviour is quite common in case of polycrystalline ceramics which is contributed by mobile charge carriers in presence of grains and grain boundaries.~\cite{kao}

\begin{figure}[t]
\centering
\includegraphics[width = 8 cm]{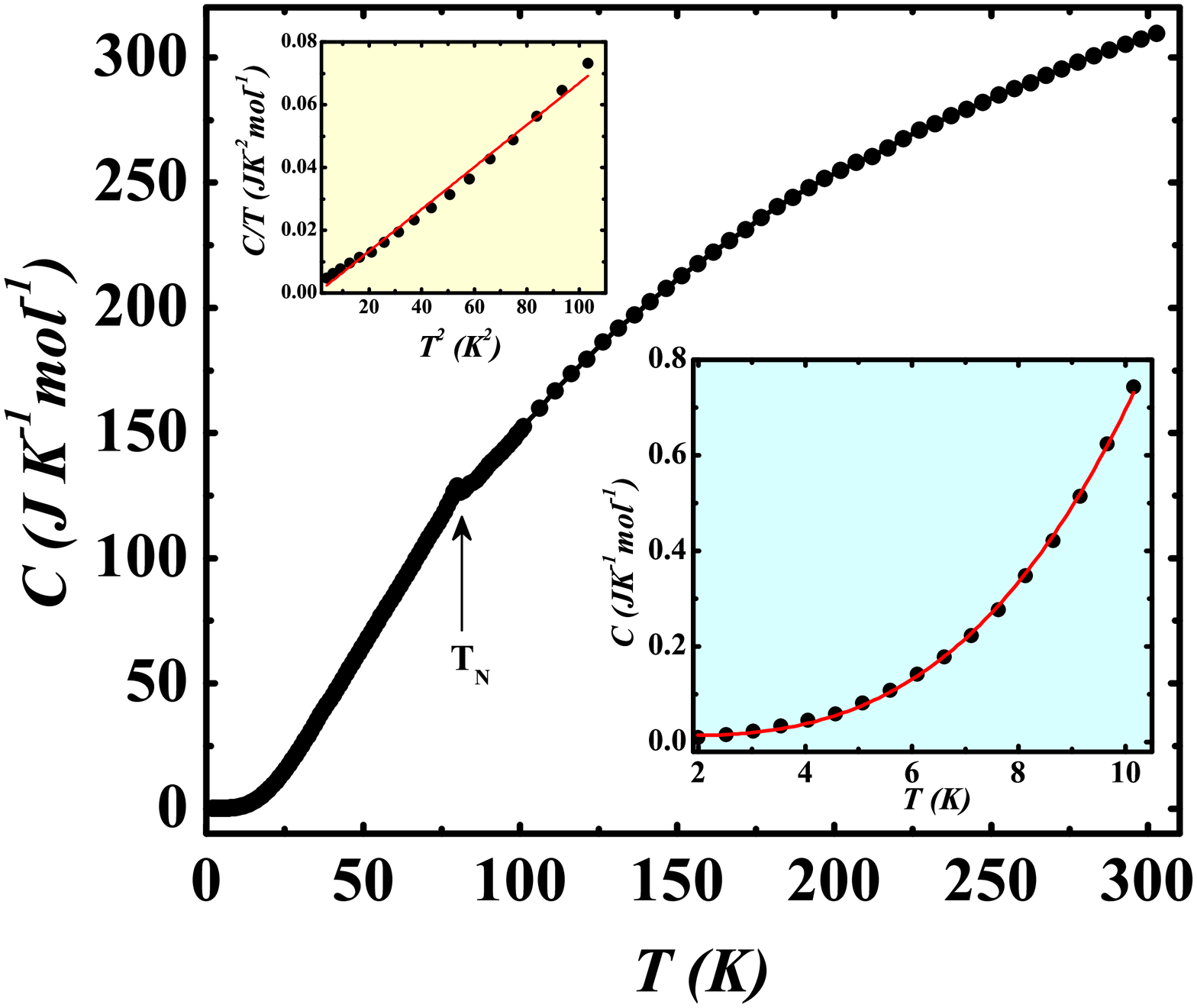}
\caption {Zero field heat capacity as a function of temperature between 2-300 K for BMO. Upper inset shows the $C/T$ versus $T^2$ plot below 10 K (solid line is a guide to the eye), while the lower inset shows the fitting to the $C(T)$ data using eqn.~\ref{clowt}. }
\label{fitting}
\end{figure}

\par
Fig. 3  displays the $T$ variation of molar heat capacity ($C$) of BMO between 2 and 300 K. A peak is observed close to the $T_N$ of the sample indicating the long range magnetic ordering. The upper inset shows the $C/T$ versus $T^2$ plot, which is nonlinear in nature. Such nonlinearlity indicates that the sample does not show a  simple $\left [\gamma T + \beta T^3\right]$ ($\gamma$ and $\beta$ are coefficients) type of heat capacity variation at low temperature (at least below 10 K). 

\section{Models and Data Analysis}
\subsection{Magnetization}
The magnetic susceptibility of a low dimensional oxide is often expressed as 
\begin{equation}
\chi = \chi_{0}+\chi_{imp}(T)+\chi_{spin}(T,J,g)
\label{chi}
\end{equation}

where $\chi_0$ is a temperature independent term arising primarily from Van-vleck paramagnetism and core diamagnetism of the sample; $\chi_{imp}$ is due to free paramagnetic impurity spins which are not in the regular lattice site; and $\chi_{spin}$ is the term arising from spin-spin correlation in a low-D (D $<$ 3) network with $J$  and $g$ are being the Heisenberg exchange integral and Land\'e-$g$ factor respectively. For the present work we have assumed $g$ = 1.9. Because, the  value of $g$ was found to be close to 1.9 for both BMO and SMO  from electron paramagnetic resonance  measurements (not shown in here) and it is kept fixed at this value for rest of  the fittings. $\chi_{imp}$ can be assumed  to vary as 
${C_{imp}}/{[T-{\theta}_{imp}]}$, where $C_{imp}$ and $\theta_{imp}$ are constants. In order to understand the magnetic state of BMO and SMO, we have  fitted the experimental $\chi(T)$ data with various theoretical models. The lower end of the fitting was restricted to 100 K due to the existence of long range magnetic order around 80 K.  
 
\subsubsection{Bonner-Fisher model for AFM spin chain}
Previously, SMO was considered as one dimensional (1D)  AFM spin-chain system  and was fitted with well known Bonner-Fisher (BF) model.~\cite{tang} We have tried to fit the $T$-dependent $\chi$ with this model.
\begin{equation}
\chi_{BF} = \frac{N\mu^2}{3k_BT}\cdot\frac{0.25 + 0.14995x +0.30094x^2}{1+1.9862x+0.68854x^2+6.0626x^3} 
\label{bf}
\end{equation}
Here $x = |J|/k_BT$,  $k_B$ is the Boltzman constant and  $N$ is the total number of spins. We have taken $N$=3$N_A$ ($N_A$ = Avogadro number) for the fitting of molar susceptibility data considering the fact that there are three Mn$^{4+}$ ions per formula unit.
The total susceptibility expression of eqn.~\ref{chi} is fitted to the experimental  $\chi_{spin} (T, J, g)$ =$\chi_{BF}$. The magnetic moment of  Mn$^{4+}$ ions is kept fixed at  $\mu = g\sqrt{S(S+1)}$ = 3.67 $\mu_B$ corresponding to $S$ = 3/2. The overall fitting by BF model is very poor (fitting for BMO is shown in the inset of fig. 4 (a)) and it cannot reproduce the peak value of the experimental $\chi(T)$ data. This indicates that the magnetic correlation in  BMO and SMO is not equivalent to 1D AFM spin-chain as speculated previously.

\subsubsection{Classical Heisenberg model for isolated spin-trimers}
Considering trimerized Mn$_3$O$_{12}$ building blocks of the compounds, we have  analyzed the data with isolated linear spin-trimer model. From the theoretical point of view this model is applicable for isolated 3-spin system where spin quantum number is large ($S \geq$ 3/2).~\cite{cregg1,cregg2,kahn,jong} The Hamiltonian for a isolated 3-spin blocks in presence of an external magnetic field  $\bf{H}$ can be written as: 
\begin{equation} 
\mathcal{H}_{t} = -J_t\sum_{i=1}^{i=2}{\bf{S_i}\cdot\bf{S_{i+1}}}-{\bf H}\cdot\sum_{i=1}^{i=3}{\bf{S_i}}
\end{equation}

Here $\bf{S_i}$ represents the $i^{th}$ spin in the trimer; and $J_t$ is the intra-trimer nearest neighbour spin-spin exchange term. The solution of the Hamiltonian gives the susceptibility from $N/3$ ($N$ is the total number of spins) independent trimers as
\begin{equation}
\chi_{t}(T)=(N/3)\dfrac{15g^2\mu_B^2}{4k_BT}\left[1+\frac{4}{3}L(\xi)+\frac{2}{3}{L(\xi)}^2\right]
\label{trimer}
\end{equation}
 
Here $\xi$ = ${15J_t}/({4k_BT})$ for $S$ = 3/2 and $L(\xi)$ is the classical Langevin function. Using  $\chi_t$ as the $\chi_{spin}$ term of eqn.~\ref{chi}, we failed to fit  the experimental data for any physically acceptable values of $J_t$ for both BMO and SMO. The fitted curve using  $\chi_t(T)$ is characteristically different from our experimental data, which shows a monotonic increase with lowering of $T$ without the signature of any hump (see the inset of fig. 4 (a)). Therefore, the magnetic interaction within the Mn$_3$O$_{12}$ polyhedra cannot be  high enough  to produce spin-trimerized state which can be held responsible for the hump like feature in the susceptibility.  
\par
We next tried to improve the model by introducing inter-trimer interaction through mean field approximation. Considering an average field $\lambda M$ experienced by the trimers due to mutual interaction, the mean field model predicts $\chi_{spin} = {\chi_t}/({1+\lambda\chi_t})$. However, incorporation of such mean field term does not improve the fitting with respect to the experimental data.

\begin{figure}[t]
\centering
\includegraphics[width = 8 cm]{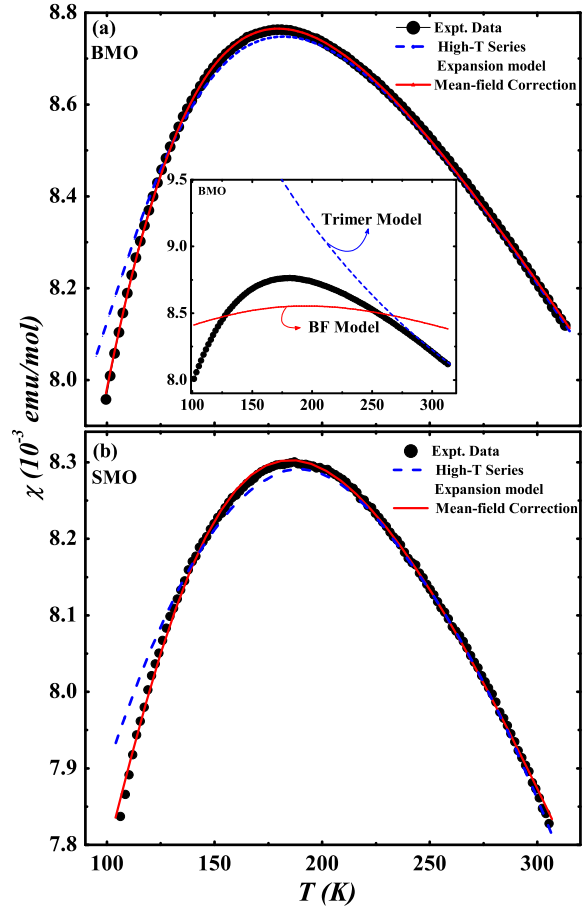}
\caption {The scattered data points in (a) and (b) respectively show the magnetic susceptibility between 100-320 K as function of temperature measured in presence of 2  kOe of magnetic field. The solid and dotted lines in the panels respectively show the fitting by pure high temperature series expansion model (eqn.~\ref{htse}) and same fitting along with mean field correction for inter-layer coupling (eqn.~\ref{mean_field}). The inset of (a) shows the fitting using Bonner-Fisher model (eqn.~\ref{bf}) and trimer model (eqn.~\ref{trimer} in case of BMO.}
\label{dielectric}
\end{figure}

\subsubsection{High-$T$ series expansion model}
It is apparent from the crystal structure  that BMO and SMO contain corrugated layers of Mn$^{4+}$ ions in the $a-c$ plane of the lattice. The Mn-Mn distance on the layer is much smaller than the Mn-Mn distance between two layers. In addition, the layers are separated from each other by non-magnetic Ba$^{2+}$ or Sr$^{2+}$ ions while Mn ions on a particular layer are connected via O$^{2-}$ ions, which can mediate superexchange. The system can have strong Mn-Mn interaction within the corrugated layer thereby paving the path for 2D magnetism. The failure of the spin-trimer model also advocates that the broad hump may be a signature of 2D magnetic correlation within the Mn layer. 

\par
For 2D Heisenberg spin system, at a sufficiently high temperature ($k_BT/J_{2D} \geq$ 0.7), the magnetic susceptibility can be expressed as a series expansion in terms of $J_{2D}/k_BT$, where $J_{2D}$ is the nearest neighbor spin spin interaction within the layer. For the studied compounds, we consider a square planar 2D lattice within the layer with the number of nearest neighbours $z_{2D}$ = 4. The magnetic interaction is expressed in terms of an isotropic $S$  = 3/2 AFM Heisenberg model with the Hamiltonian: $\mathcal{H}_{2D}=\sum_{nn}J_{2D}\bf{S_i}\cdot\bf{S_j}$, where the  summation runs over all pair of nearest neighbour spins.~\cite{lines} Considering high-$T$ series expansion~\cite{rushbrooke,lines, navarro, yamaji} up to sixth order term, the spin susceptibility can be expressed as

\begin{equation}
\chi_{2D}=\frac{Ng^2{\mu_B}^2}{J_{2D}(3\Theta+\sum_{n=1}^6\frac{a_n}{\Theta^{n-1}})}
\label{htse}
\end{equation}

Here $\Theta = {4k_BT}/{15J_{2D}}$ for $S$ = 3/2, and  the values of $a_n$ are  4.0,~1.6,~0.304,~0.249,~0.132 and 0.013 respectively. We have tried to fit the experimental susceptibility between temperature 100$-$320 K by incorporating the above expression of $\chi_{2D}$  for $\chi_{spin}$ in  eqn.~\ref{chi}.  For BMO and SMO, the best fittings  for this 2D high temperature series expansion model are obtained for $J_{2D}\approx -$45 K and $-$47 K respectively. From fig. 4 (a) and (b) it is evident that the fittings are reasonably good above 125 K. $\chi_{imp}$ contribution is found to be quite small (for example,  $C_{imp}$ = 0.039(1) emu.K.mol$^{-1}$ and $\theta_{imp}$ =$-$9.3(2) K for BMO) and these value are reasonable considering small impurity contribution in the samples.

\par
 For 2D Heisenberg model, the susceptibility maximum  is given by ${k_BT_{{max}}}/{|J_{2D}|}=1.12S(S+1)+0.10$. For BMO with $S$ = 3/2 and $T_{max}$ =179 K, we obtain $|J_{2D}|$ = 42 K, which is quite close to the value obtained by fitting the full curve with high temperature expansion model. This once again ensures the applicability of 2D model to these layered magnetic materials. 

\subsubsection{Modified High-$T$ series expansion with mean field type inter-layer interaction}
It is clear from fig. 4 (a) and (b) that the high-$T$ series expansion deviates from the experimental data below about 125 K. A possible reason for such deviation is the existence of inter-layer magnetic interaction. If the inter-layer coupling is considered in terms of a molecular field type model, the total susceptibility can be expressed as,
\begin{equation}
\chi_{3D} = \frac{\chi_{2D}}{ 1-\left(2J^{\prime}z^{\prime}(k_BT)^2/Ng^2\right)\chi_{2D}}
\label{mean_field}
\end{equation}

Here $J^{\prime}$ and $z^{\prime}$ are  the inter-layer exchange and coordination number respectively.~\cite{ginsberg} We have fitted the susceptibility data using above expression of $\chi_{3D}$ for $\chi_{spin}$ in eqn.~\ref{chi} and the fittings are shown in fig. 4 (a) and (b). It is  evident that the introduction of this mean field model improves the fitting and we can  fit the data over a wider range of temperature. We take $z^{\prime}$ = 2 for the present case (since each Mn has two inter-layer nearest neighbours), and for BMO and SMO, the best fit is obtained for $J_{2D}$ and $J^{\prime}$ values mentioned in table~\ref{values}. The negative sign of $J^{\prime}$ indicates AFM inter-layer interaction, which is  supported by our isothermal $M$ versus $H$ data.  

\subsection{Heat Capacity}
The total heat capacity contains both lattice ($C_{latt}$)  and magnetic ($C_{mag}$) parts and the magnetic part is generally estimated by subtracting $C_{latt}$ from total $C$. However, due to  nonavailability of a proper nonmagnetic counterpart, it is quite difficult to have a prior knowledge of $C_{latt}$. We therefore only looked at the low temperature part of $C$ (below 10 K), where $C_{latt}$ is quite small and generally follows a Debye $T^3$ law. It is already evident that $C/T$ versus $T^2$ below 10 K does not follow a linear trend. In general (ignoring the electronic part as the sample is an insulator) at low temperature ($T < \Theta_{Debye}$/10, $\Theta_{Debye}$ = Debye temperature) for a magnetic system, $C = \beta T^3 + C_{mag}$. The spin wave contribution of $C$ for a magnetically ordered system at sufficiently low temperature can be expressed as $C_{mag} \sim T^{d/n}$, where $d$ is the effective dimension of magnetic interaction and $n$ = 2 and 1 respectively for FM and AFM orderings~\cite{jong}. For the present system, the magnetic state is AFM, and therefore we have taken $n$ = 1.  For a 3D AFM system (with $d$= 3), $C_{mag}$ is expected to show $T^{3}$ nature, which is inseparable from the lattice part. However, since we get a nonlinear $C/T$ versus $T^2$ curve, a simple $T^3$ nature of $C_{mag}$ is unacceptable. We have plotted $C/T^2$ versus $T$, and it gives a better linearity than the $C/T$ versus $T^2$ plot. This indicates that a predominant $T^2$ term is present in the low-$T$ $C(T)$ data, which can originate from a 2D AFM ordering ($d$ = 2, $n$ = 1) with $C_{mag} = \alpha T^2$.

However, the $C$ versus $T$ data can be  much better fitted  if we incorporate an addition term,  $C_{hf} = \delta T^{-2}$, resulting,

\begin{equation}
C = \beta T^3 + \alpha T^2 + \delta T^{-2}
\label{clowt}
\end{equation} 

It is to be noted that $C_{hf}$ term  is found to exist in several transition metal oxides including perovskite based manganites~\cite{wood, martin} and it is attributed to the nuclear hyperfine contribution to $C$. If the atomic nucleus has a magnetic moment $\mu_I$, there can be hyperfine splitting of the energy level due to the effective magnetic field ($H_{hf}$) produced at the nuclear site by the electrons in the partially filled shells and thermal excitation across these hyperfine levels contributes towards $C_{hf}$. The fitting of $C(T)$ by the combination of three terms (as in eqn.~\ref{clowt}) is shown in the lower inset of fig. 3 along with the experimental data. The values of the fitted parameters are found to be $\beta$ = 0.47(3) mJ.mol$^{ -1}$.K$^{-4}$, $\alpha$ = 4.71(8) mJ.mol$^{-1}$.K$^{-3}$ and $\delta$ = 19.98(4) mJ.K.mol$^{-1}$.        
\par
If we assume that the $T^3$ contribution is solely arising from the lattice part, one can write $\Theta_{Debye}^3 = {12\pi^4pR}/{5\beta}$, where $p$ is the number of atoms per formula unit. Using the fitted value of $\beta$, we obtain $\Theta_{Debye}$ = 412 K. The coefficient of the 2D spin wave term $\alpha$ has a value comparable to many other layered magnetic systems including manganites.~\cite{wood, yang}

\par
One can calculate the effective hyperfine field from the value of  $\delta$ using,

\begin{equation}
\delta = \frac{Rp}{3}\left(\frac{I+1}{I}\right)\left(\frac{\mu_IH_{hf}}{\mu_0 k_B}\right)^2 
\label{hyperfine}
\end{equation} 
 Here $I$ is the nuclear spin, $\mu_I$ is the nuclear magnetic moment and $\mu_0$ is the free space permeability. The hyperfine interaction is only important for Mn nuclei, as it has only partially filled electronic shell to contribute towards hyperfine field at the nuclear site. For Mn, $I$ = 5/2 and using this value in eqn.~\ref{hyperfine} we get $H_{hf} \approx$  9 $\times$10$^7$ A/m which is equivalent to 2 $\times$10$^6$ Oe (in cgs). This  value is comparable to the hyperfine field in other manganites.~\cite{martin} 
  
\section{Summary and Conclusion}

The present work brings out several interesting aspects regarding the corrugated layered manganites BMO and SMO. Despite the fact that the samples contain Mn$_3$O$_{12}$ building blocks, models based on independent as well as interacting spin-trimers (interaction taken into account via mean field) can not interpret the susceptibility data. The broad peak in the $\chi(T)$ data, a typical signature of low-D magnetic correlation, rather found to be associated with 2D magnetic interaction within the corrugated layers. The high-$T$ series expansion of Heisenberg model turns out  to be appropriated for the interpretation of susceptibility data. The calculated values of the intra-layer coupling term $J_{2D}$ are consistent with the maximum in $\chi(T)$ for both the samples. 
\par
An additional inter-layer magnetic interaction term $J^{\prime}$ is required to fit the $\chi(T)$ data below about 125 K for both BMO and SMO. The intra-layer Mn-Mn distance is about 2.5 \AA, while the inter-layer distance is about 5.7 \AA. Therefore, there is a chance for the existence of finite $J^{\prime}$ in the samples. The ratio $\left({J^{\prime}}/{J}\right)$ $\approx$ 0.05 which is close to the values found in other layered magnetic systems. It is to be noted that the value of $J_{2D}$ is slightly higher in SMO than BMO (see table I), which is consistent with the fact that the Mn-Mn separation on a particular layer is  slightly shorter in the former composition. Similarly,  $T_N$ is  slightly higher in SMO possibly due to the same reason.   

\par
The 2D nature of the magnetic correlation is further evident from the low-$T$ (below 10 K) heat capacity data of BMO where a clear existence of $T^2$ term corresponding to the 2D AFM spin wave excitation is found. This is possibly due to the fact that the magnetic ordering in the compound occurring around $T_N$ = 80 K has 2D character. It is difficult to rule out the possibility of a long range 3D AFM ordering completely from the present $C(T)$, because a 3D AFM state would give rise to a $T^3$ variation which is inseparable from the lattice part. However, it is true that the  sample indeed retains 2D spin wave excitation down to the lowest temperature which is likely to be connected with the layered crystal structure. 

\par
An intriguing aspect of the present work is the observation of peak like dielectric anomaly close to the AFM ordering temperature in both the studied materials. This is quite an important observation considering the recent interest in multiferroic oxides.~\cite{cheong} Dielectric anomaly does not show any relaxation with frequency indicating that it is associated with some electric ordering. However, we failed to observe any spontaneous electric polarization below $T_N$ indicating that the ordering is not ferroelectric type. Possibly, the sample undergo an antiferroelectric (AFE) order around $T_N$ which does not generate macroscopic $P$. Similar AFM/AFE type multiferroicity has been observed in several other systems such as  Dy$_3$Fe$_5$O$_{12}$ or LiCrO$_2$.~\cite{rogers,licro2} 

\par
In conclusion, we observe that the compounds Ba$_4$Mn$_3$O$_{10}$ and Sr$_4$Mn$_3$O$_{10}$ show strong 2D magnetic character as opposed to the spin-chain like 1D character reported earlier.~\cite{tang} The low temperature heat capacity in Ba$_4$Mn$_3$O$_{10}$ indicates the presence of 2D spin wave excitation. Interestingly, both the compounds show dielectric anomaly devoid of any spontaneous electric polarization around $T_N$, and it might be an indication of the onset of AFE state at the magnetic transition point.

\section{Acknowledgment}
{\it Unit of Nanoscience at IACS} is duly acknowledged for magnetic measurements. Authors would like to  thank CSIR, India for financial support (grant number: 03(1209)/12/EMR-II).  DST, India is also acknowledged for low temperature and high magnetic field  facilities at UGC-DAE Consortium for Scientific Research at Indore center.


\begin{thebibliography}{99}

\bibitem{salamon} M. B. Salamon, Rev. Mod. Phys. {\bf73}, 583 (2001). 

\bibitem{camno} A. I. Mihut, L. E. Spring, R. I. Bewley, S. J. Blundell, W. Hayes, Th. Jest\"adt, B. W. Lovett, R. McDonald, F. L. Pratt, J. Singleton, P. D. Battle, J. Lago, M. J. Rosseinsky, and J. F. Vente, J. Phys.: Condens. Matter {\bf10}, L727 (1998).

\bibitem{srmno} N. Floros, M. Hervieu, G. van Tendeloo, C. Michel,
A. Maignan, and B. Raveau, Solid State Sci. {\bf 2}, 1 (2000).

\bibitem{bamno}V. G. Zubkov, A. P. Tyutyunnik, I. F. Berger,V. I. Voronin, G. V. Bazuev,
C. A. Moore, P. D. Battl, J. Solid State Chem. {\bf 167}, 453 (2002) .

\bibitem{kb} K. Boulahya, M. Parras, J. M. Gonz\'alez-Calbet, U. Amador, J. L. Mart\'inez, and M. T. Fern\'andez-D\'iaz, Phys. Rev. B {\bf 69}, 024418 (2004).

\bibitem{tang} Y. Tang, X. Ma, Z. Kou, Y. Sun, N. Di, Z. Cheng, and Q. Li, Phys. Rev. B {\bf 72}, 132403 (2005).

\bibitem{bf} J. C. Bonner and M. E. Fisher, Phys. Rev {\bf 135}, A640 (1964).

\bibitem{boulahya} K. Boulahya, M. Parras, U. Amador, and J. M. Gonz\'{a}lez-Calbet, Solid State Ionics {\bf172}, 543 (2004).

\bibitem{chattopadhyay} S. Chattopadhyay, S. Giri, and S. Majumdar, J.Phys.: Condens. Matter {\bf23}, 216006 (2011).

\bibitem{hvr} H. V. Rietveld, J. Appl. Crystaloogr. {\bf 2}, 65 (1969).

\bibitem{maud} http://www.ing.unitn.it/$\sim$maud/

\bibitem{kao} K. C. Kao, Dielectric Phenomena in Solids With Emphasis on Physical Concepts of Electronic Processes,(Elsevier Academic Press, California, USA, 2004). 

\bibitem{cregg1} P. J. Cregg, J. L. Palacios, P. Svedlindh, and K. Murphy, J.Phys.: Condens. Matter {\bf20}, 204119 (2008).

\bibitem{cregg2} P. J. Cregg, J. L. Palacios, and P. Svedlindh, J. Phys. A: Math. Theor. {\bf41}, 435202 (2008).

\bibitem{kahn} O. Kahn, Molecular Magnetism (VCH Publishers Inc., 1993).

\bibitem{jong} L. J. de Jongh and A. R.Miedema, Adv. Phys. {\bf 23}, 1 (1974).

\bibitem{lines} M. E. Lines, J. Phys. Chem. Solids {\bf31}, 101 (1970).

\bibitem{rushbrooke} G. S. Rushbrooke, and P. J. Wood, Molec. Phys. {\bf1}, 257 (1958).

\bibitem{yamaji} K. Yamayaji, and J. Kondo, J. Phys. Soc. Jpn. {\bf 35}, 25 (1973).

\bibitem{navarro} R. Navarro, Magnetic Properties of Layered Transition Metal Compounds, ed. L. J. de Jongh (Kluwar Academic Publisher, 1990), p. 105-190. 

\bibitem{ginsberg} A. P. Ginsberg, and M. E. Lines, Inorg. Chem. {\bf 11}, 2289 (1972).

\bibitem{wood} B. F. Woodfield, M. L. Wilson, and J. M. Byers, Phys. Rev. Lett. {\bf 78}, 3201 (1997).

\bibitem{martin} M. R. Lees, O. A. Petrenko, G. Balakrishnan, and D. McK. Paul, Phys. Rev. B  {\bf 59}, 1298 (1999).

\bibitem{yang} H.D. Yang, I.P. Hong, F.H. Hsu, H.H. Li, S.Y. Tu, H.L. Huang, S. Chatterjee, S. Mollah, Y.-K. Kuo, T.I. Hsu, H.C. Ku, and W.-H. Li, Solid State Commun. {\bf 27}, 229 (2003).

\bibitem{cheong} S. – W. Cheong, and M. Mostovoy, Nat. Mater. {\bf 6}, 13 (2007).

\bibitem{rogers} P. D. Rogers, Y. J. Choi, E. C. Standard, T. D. Kang, K. H. Ahn, A. Dubroka, P. Marsik, Ch. Wang, C. Bernhard,S. Park, S.-W. Cheong, M. Kotelyanskii, and A. A. Sirenko, Phys. Rev. B {\bf 83}, 174407 (2011).

\bibitem{licro2} S. Seki, Y. Onose, and Y. Tokura, Phys. Rev. Lett. {\bf 101}, 067204 (2008)

\end{thebibliography}
\end{document}